\documentclass[a4paper,11pt]{article}

\usepackage[round,authoryear,sort]{natbib}
\bibpunct{(}{)}{,}{a}{,}{,~}

\title{Why special relativity should not be a template for a fundamental reformulation of quantum mechanics}

\author{ Harvey R. Brown\\
Faculty of Philosophy, University of Oxford\\ 10 Merton Street, Oxford OX1
4JJ, U.K.\\{\em harvey.brown@philosophy.ox.ac.uk}
\and Christopher G. Timpson\\ Division of History and Philosophy of Science, School of Philosophy\\ University of Leeds, LS2 9JT, UK.\\ {\em c.g.timpson@leeds.ac.uk}}

\date{July 2005}

\begin{document}

\maketitle

\noindent\textit{The principle of relativity is a principle that narrows the possibilities; it is not a model, just as the second law of thermodynamics is not a model.} Albert Einstein\footnote{This statement was made by Einstein in 1911 at a scientific meeting in Zurich; see \citet{galison04}, p. 268. In 1911 Einstein was still using ``principle of relativity" to mean theory of relativity.}
\bigskip

\begin{abstract}
\noindent In a comparison of the principles of special relativity and of quantum mechanics, the former theory is marked by its relative economy and apparent explanatory simplicity. A number of theorists have thus been led to search for a small number of postulates---essentially information theoretic in nature---that would play the role in quantum mechanics that the relativity principle and the light postulate jointly play in Einstein's 1905 special relativity theory. The purpose of the present paper is to resist this idea, at least in so far as it is supposed to reveal the fundamental form of the theory. It is argued that the methodology of Einstein's 1905 theory represents a victory of pragmatism over explanatory depth; and that its adoption only made sense in the context of the chaotic state state of physics at the start of the 20th century---as Einstein well knew.
\end{abstract}\

\section{Quantum mechanics: the CBH theorem}

In an important recent development in quantum mechanics, Clifton, Bub and Halvorson (henceforth CBH) have shown that the observables and state space of a physical theory must be quantum mechanical if three `information-theoretic' constraints hold.\footnote{\citet{cliftonetal03}.} The constraints are:
\begin{enumerate}
\item no superluminal information transmission between two systems by measurement on one of them,
\item no broadcasting of information contained in an unknown physical state,\footnote{The notion of broadcasting, and the fact of its failure in quantum theory, is due to \citet{nobroadcasting}. No-broadcasting is a more distinctively quantum feature than the more familiar no-cloning theorem.} and
\item no unconditionally secure bit-commitment.
\end{enumerate}
The CBH theorem states that these constraints force any theory formulated in $C^\ast$-algebraic terms to incorporate a non-commuting algebra of observables for individual systems, kinematic independence for the algebras of space-like separated systems and the possibility of entanglement between space-like separated systems. (Conversely, any $C^\ast$-algebraic theory with these distinctively quantum properties will satisfy at least the three information-theoretic constraints.\footnote{CBH showed ({\em op. cit}) that such quantum properties imply the first two constraints, and \citet{halvorson04} showed that the third constraint related to bit-commitment also follows.})

This result is not only of great interest in itself, but it appeared at a time when attention to the putatively fundamental role that the notion of information plays in understanding quantum theory has been growing significantly.  It is not our aim in this paper to examine in detail either the scope of the theorem\footnote{In this connection see \citet{valentini03} and \citet{timpson04}, section 9.2.2. We feel it worthwhile pointing out that in non-relativistic quantum mechanics, it has long been accepted that signalling at infinite speeds is a theoretical possibility. For example, a particle strictly confined to a region of compact support by means of a potential barrier can propagate to arbitrary distances in arbitrarily short times when the barrier is suddenly removed. This does not violate the no-signalling theorem in quantum mechanics because the latter is defined with respect to communication between pairs of entangled systems. But what this case emphasizes is that the no-superluminal-information-transmission constraint in the CBH theorem is of limited validity, at least in non-relativistic quantum mechanics.}, or the contentious issue of the role of information in modern physics\footnote{See \citet{timpson02,timpson03,timpson04}.}.  We are concerned with the methodological issues at stake. At the start of their paper, CBH wrote:

\begin{quotation}
\begin{sloppypar}
\noindent The fact that one can characterize quantum theory \ldots in terms of just a few simple information-theoretic principles \ldots lends credence to the idea that an information-theoretic point of view is the right perspective to adopt in relation to quantum theory. Notice, in particular, that our derivation links information-theoretic principles directly to the very features of quantum theory---noncommutativity and nonlocality---that are so conceptually problematic from a purely physical/mechanical point of view.
We therefore suggest substituting for the conceptually problematic mechanical perspective on quantum theory an information-theoretic perspective. That is, we are suggesting that quantum theory be viewed, not as first and foremost a mechanical theory of waves and particles \ldots but as a theory about the possibilities and impossibilities of information transfer.\footnote{\citet{cliftonetal03}, p. 4.}
\end{sloppypar}
\end{quotation}

Even more significantly for our purposes, at the end of their paper CBH suggested an analogy between their characterization of quantum mechanics and Albert Einstein's special theory of relativity (henceforth SR). The ``foundational significance'' of the CBH derivation is, according to these authors, that quantum mechanics should be interpreted as a {\em principle theory}, in the sense of the term that Einstein used to describe his 1905 formulation of SR.\footnote{\citet{cliftonetal03}, p. 24.}. CBH saw their constraints as analogous to the principles---the relativity principle and the light postulate---used by Einstein to derive the nature of relativistic kinematics.

There can be no doubt that Einstein's 1905 treatment of relativistic kinematics was a triumph of economy in relation to the corresponding treatment of moving rods and (to the extent it existed, as we see below) clocks provided by the leading {\em fin de si\`{e}cle} ether theorists. But it is still not sufficiently appreciated that by his own admission, Einstein's principle theory route was based on a policy of despair, and represented a strategic retreat from the more desirable but, in his view, temporarily unavailable {\em constructive} approach. It is worth dwelling a little on this historical episode, to see what implications it might have for the CBH program.\footnote{The present paper, sections 2, 3 and 4 of which draw heavily on \citet{brown05a,brown05b}, is a development of views expressed in \citet{timpson04}, section 9.2.}

\section{Special relativity as a ``principle theory''}

It is well known that the principle/constructive theory distinction was articulated by Einstein in a popular article on his theory of relativity published in 1919 in the London \emph{Times}\footnote{\citet{einstein19}.}. But it was a theme that appeared sporadically throughout his life-long writings.

In January 1908, roughly two and a half years after publishing his celebrated paper on special relativity\footnote{\citet{einstein05b}.}, Einstein wrote in a letter to Arnold Sommerfeld:
\begin{quote}So, first to the question of whether I consider the relativistic treatment of , e.g., the mechanics of electrons as definitive. No, certainly not. It seems to me too that a physical theory can be satisfactory only when it builds up its structures from {\em elementary} foundations. The theory of relativity is not more conclusively and absolutely satisfactory than, for example, classical
thermodynamics was before Boltzmann had interpreted entropy as
probability. If the Michelson-Morley experiment had not put us in
the worst predicament, no one would have perceived the relativity
theory as a (half) salvation.\footnote{\citet{einstein95a}.}
\end{quote}
Einstein is repeating here an analogy between SR and thermodynamics that he had mentioned in a published note addressed to Ehrenfest already in 1907, in which he compared SR with ``the second law of the theory of heat.''\footnote{\citet{einstein07a}.} In both cases, Einstein was emphasizing the {\em limitations} of SR, not its strengths.

In order to see why SR is only a `half' salvation, consider for a minute the analogy with thermodynamics.

Think of an idealized single-piston heat engine undergoing a Carnot cycle, and consider the theoretical limits of its efficiency. Such limits can in principle be established by exploiting knowledge of the micro-structure of the working substance of the engine, and in particular by using the principles of statistical mechanics that apply to the molecular structure of the gas in the piston. A much easier approach, however, is to fall back on the laws
of classical thermodynamics to shed light on the performance of the
engine---phenomenological laws which stipulate nothing about the deep structure of the working substance. According to this approach, the efficiency of the heat engine must depend in a certain way on the ratio of the temperatures of the two heat reservoirs simply because, whatever the gas in the piston is made up of, if it did not it would be possible for the engine to act as a perpetual motion machine of 'the second kind'. And this possibility is simply ruled out by hypothesis in thermodynamics. 

Yet it is hard to not to wonder why, after all, such a perpetual motion cannot exist. Indeed, it is widely held that statistical mechanics in principle explains why (even if the details involved are controversial). But thermodynamics cannot. The impossibility of perpetual motion machines of various kinds is the very starting point of thermodynamics. What this theory gains in practicality and in the evident empirical solidity of its premisses, it loses in providing physical insight.

Einstein considered thermodynamics as the archetypical example of what he would call in 1919 a principle theory in physics, one which is based on well verified, but unexplained observable regularities. On the other hand, statistical mechanics, or more specifically the kinetic theory of gases, was for Einstein the prime example of a constructive theory, one built on the ``elementary foundations'' mentioned in his 1908 letter. These foundations involve hypotheses about unseen fundamental processes---normally involving the microstructure of bodies and its mechanical principles. 

The distinction has been the subject of increasing attention in recent years\footnote{See, for example, \citet{brownpooley01,brownpooley06} and \citet{balashovjanssen03}.}, but it is easily misunderstood. First, it is clearly not categorical: all theories have principles, it is just that some are more phenomenological than others. Thermodynamics and statistical mechanics are on opposite ends of a spectrum of possible theories, and there are indeed respectable theories---as we shall see below---which lie somewhere in between. 

Principle theories are typically employed when constructive theories are either unavailable,  too difficult to build, or relatively unwieldy. For according to Einstein, ``when we say we have succeeded in understanding a group of natural processes, we invariably mean that a constructive theory has been found which covers the processes in question.''\footnote{\citet{einstein19}.} Yet, Einstein stressed that SR is a principle theory. Why then did he feel it necessary to sacrifice explanatory content in developing his theory of relativity?

\section{Rods, clocks, and the quantum}

Recall the title of Einstein's 1905 relativity paper: ``On the electrodynamics of moving bodies''. One of the great challenges of late nineteenth century electrodynamics and optics was to predict the outcome of experiments involving electromagnetic phenomena being performed in a laboratory {\em moving with respect to the luminiferous ether}. After all, the earth is in motion relative to the centre of mass of the solar system, and at least some of the time must be moving relative to the ether---the invisible seat of electromagnetic phenomena. But by the turn of the century, the ether had become in the minds of some experts a very shadowy entity indeed. Made of an obscure kind of  ``imponderable matter'', its main role was increasingly just that of providing the inertial frame of reference relative to which the fundamental electromagnetic field equations of Maxwell were postulated to hold. The question was now: what form do the field equations have in earth-bound frames that are moving relative to this fundamental frame?

Einstein is famous for claiming in 1905, on the basis of his relativity principle, that all laws of physics, including those of electrodynamics, take the same form in all inertial reference frames, so happily Maxwell's equations can be used just as well in the moving laboratory frame. But this conclusion, or something very close to it, had already been anticipated by several great ether theorists, including H. A. Lorentz, Joseph Larmor and particularly Henri Poincar\'{e}. This was largely because there had been from the middle of the nineteenth century all the way to 1905 a series of experiments involving optical and electromagnetic effects that failed to show any sign of the ether wind rushing through the laboratory: it was indeed as if the earth was always at rest relative to the ether. (The most famous of these, and the most surprising,  was of course the 1887 Michelson-Morley experiment.) Like the above-mentioned ether theorists, Einstein realized  that the covariance of Maxwell's equations---the form invariance of the equations---is achieved when the relevant coordinate transformations take a very special form, but Einstein was unique in his understanding that these transformations, properly understood, encode new predictions as to the behaviour of rigid bodies and clocks in motion. That is why, in Einstein's mind, a new understanding of space and time themselves was in the offing. 

Both the mathematical form of the transformations, and at least the non-classical distortion of moving rigid bodies were already known to Lorentz, Larmor and Poincar\'{e}---indeed a family of possible deformation effects was originally suggested independently  by Lorentz and G. F. FitzGerald to explain the Michelson-Morley result.\footnote{For recent treatments of this episode, see \citet{brown01,brown05b}.} It was the connection between them, i.e. between the coordinate transformations and motion-induced deformation, that had not been fully appreciated before Einstein. In the first (``kinematical'') part of his 1905 relativity paper, Einstein established the operational meaning of the so-called Lorentz coordinate transformations and showed that they lead not just to a special case of FitzGerald-Lorentz deformation (longitudinal contraction), but also to the ``slowing down'' of clocks in motion---the phenomenon of time dilation. Now it is still not well known that Larmor and Lorentz had come tantalizingly close to predicting this phenomenon; they had independently seen just before the turn of the century how it must hold in certain very special cases. But as a general effect that does not depend on the constitution of a clock, its discovery was Einstein's own. 

Einstein did something else that was new and important in the kinematical part of his paper. He derived the Lorentz transformations not from the symmetry properties of Maxwell's equations, but by using an argument inspired by thermodynamics. The reason lies in his earlier investigations of the properties of black-body radiation.

Several months before he wrote his paper on SR, Einstein had written a revolutionary paper claiming that electromagnetic radiation has a granular structure. The suggestion that radiation was made of quanta---or photons as they would later be dubbed---was the basis of Einstein's extraordinary treatment of the photoelectric effect in the same paper. But the immediate consequence of Einstein's commitment to the photon was to destabilize in his mind all the previous work on the electrodynamics of moving bodies.

All the work of the ether theorists was based on the assumption that Maxwellian electrodynamics is strictly true, and not just true on average. In the work of Lorentz, Larmor and Poincar\'{e}, the Lorentz transformations make their appearance as symmetry transformations (whether considered approximate or otherwise) of these equations. But Maxwell's equations are incompatible with the existence of the photon.

In his 1949 {\em Autobiographical Notes}, published when he was 67, Einstein was clear about the seismic implications of this conundrum.
\begin{quotation}
\noindent Reflections of this type [on the dual wave-particle nature of radiation] made it clear to me as long ago as shortly after 1900, i.e., shortly after Planck's trailblazing work, that neither mechanics nor electrodynamics could (except in limiting cases) claim exact validity. By and by I despaired of the possibility of discovering the true laws by means of constructive efforts based on known facts.\footnote{\citet{einstein69}, p. 51, 53}
\end{quotation}
Already in the {\em Notes}, Einstein had pointed out that the general validity of Newtonian mechanics came to grief with the success of the electrodynamics of Faraday and Maxwell, which led to Hertz's detection of electromagnetic waves---``phenomena which by their very nature are detached from every ponderable matter''.\footnote{{\em Op. cit.}, p. 25} Later, he summarized the nature of Planck's 1900 derivation of his celebrated black-body radiation formula, in which quantization of absorption and emission of energy by the mechanical resonators is presupposed. Einstein noted that although this contradicted the received view, it was not immediately clear that electrodynamics---as opposed to mechanics---was violated. But now with the emergence of the light quantum, not even electrodynamics was sacrosanct. 
\begin{quote}
All my attempts \ldots to adapt the theoretical foundation of physics to this [new type of] knowledge failed completely. It was if the ground had been pulled out from under one, with no firm foundation to be seen anywhere, upon which one could have built.\footnote{{\em Op. cit.}, p. 45.}
\end{quote}

Earlier in the {\em Notes}, Einstein had sung the praises of classical thermodynamics, ``the only physical theory of universal content concerning which I am convinced that, within the framework of the applicability of its basic concepts, it will never be overthrown''. Now, he explains how the very structure of the theory was influential in the search for a way out of the turn-of-the-century crisis in physics.
\begin{quote}
The longer and more despairingly I tried, the more I came to the conviction that only the discovery of a universal formal principle could lead us to assured results. The example I saw before me was thermodynamics. The general principle was there given in the theorem\footnote{The word ``theorem'' for ``Satze'' in the translation by P. A. Schilpp  is perhaps better rendered as ``sentence'' or ``statement''. One of us (H.R.B.) thanks Thomas M\"{u}ller for discussion of this point.}: the laws of nature are such that it is impossible to construct a {\em perpetuum mobile} (of the first and second kind). How, then, could such a  universal principle be found?\footnote{{\em Op. cit.}, p. 53.} 
\end{quote}

\section{Einstein's doubts}

It is well-known that Einstein's based his derivation of the Lorentz transformations on a combination of the relativity principle (essentially the same as that defended by Newton) and his so-called light postulate. (The latter was the claim that relative to a certain inertial frame, the speed of light is independent of the speed of the source and isotropic---something every ether theorist took for granted when the frame in question is taken to be the fundamental ether rest frame\footnote{In 1921, Wolfgang Pauli would correctly describe Einstein's light postulate as the  ``true essence of the old aether point of view'';  \citet{pauli81}, p. 5. It should also be noted that the derivation of the Lorentz transformations requires a third, admittedly innocuous, assumption: the isotropy of space.} and something which remarkably Einstein felt would survive whatever the eventual quantum theory of radiation would reveal.) He showed that length contraction for rigid rods and time dilation for ideal clocks are consequences of these phenomenological assumptions, in the same way that, say, the existence of entropy and its non-decreasing behaviour over time for adiabatic systems are a consequence of the laws of thermodynamics. Of course, the precise form of the phenomena of contraction and dilation depended on Einstein's choice of a convention for spreading time through space in both the resting and moving frames---a choice Poincar\'{e} had already advocated.

Einstein would have preferred a constructive account of these relativistic effects, presumably based on the nature of the non-gravitational forces that hold the constituent parts of rods and clocks together. But as we have seen, for Einstein the elements of such an account were not to be had in 1905. The price to be paid for the resulting strategic retreat to a principle theory approach was not just loss of insight; Einstein became increasingly uneasy about the role played by rods and clocks in this approach. This unease is seen in a paper entitled ``Geometry and Experience'' he published in 1921\footnote{\citet{einstein21}.}, and in particular in his  1949 \emph{Autobiographical Notes}:

\begin{quotation}
  One is struck [by the fact] that the theory [of special relativity]
  \ldots{} introduces two kinds of physical things, i.e., (1)
  measuring rods and clocks, (2) all other things, e.g., the
  electromagnetic field, the material point, etc. This, in a certain
  sense, is inconsistent; strictly speaking measuring rods and clocks
  would have to be represented as solutions of the basic equations
  (objects consisting of moving atomic configurations), not, as it
  were, as theoretically self-sufficient entities. However, the
  procedure justifies itself because it was clear from the very
  beginning that the postulates of the theory are not strong enough to
  deduce from them sufficiently complete equations \ldots{} in order
  to base upon such a foundation a theory of measuring rods and
  clocks. \ldots{} But one must not legalize the mentioned sin so far
  as to imagine that intervals are physical entities of a special
  type, intrinsically different from other variables (`reducing
  physics to geometry', etc.).\footnote{\citet{einstein69}, pp. 59, 61.}
\end{quotation}
These remarks are noteworthy for several reasons.

First, there is the issue of justifying the ``sin'' of treating rods and clocks as primitive, or unstructured entities in SR. Einstein does not say in 1949, as he did in 1908 and 1921, that the ``elementary'' foundations of a constructive theory of matter are still unavailable; rather he simply reminds us of the limits built into the very form of the 1905 theory. It is hardly any justification at all. Considerable progress in the relativistic quantum theory of matter {\em had} been made between 1905 and 1949. Was it Einstein's long-standing distrust of the quantum theory that held him back from recognizing this progress and its implications for his formulation of SR?

Second, consider the criticism Abraham Pais made of H. A. Lorentz in his acclaimed 1982 biography of Einstein: ``Lorentz never fully made the transition from the old dynamics to the new kinematics.''\footnote{\citet{pais82}, p. 167.} As late as 1915 Lorentz thought that the relativistic contraction of bodies in motion can be explained if the known property of distortion of the electrostatic field surrounding a moving charge is supposed to obtain for all the other forces that are involved in the cohesion of matter. In other words, Lorentz viewed such kinematical effects as length contraction as having a dynamical origin, and it is this notion that Pais found reprehensible. Yet, when Einstein appeals to the nature of rods and clocks as ``moving atomic configurations'', it seems that not even he ever fully accepted the distinction between dynamics and kinematics. For to say that length contraction is intrinsically kinematical would be like saying that energy or entropy are intrinsically thermodynamical, not mechanical---something Einstein would never have accepted.\footnote{Joseph Larmor commented in relation to Einstein's 1905 relativity paper that it actually contained dynamical reasoning  ``masquerading in the language of kinematics''; \citet{larmor29}, p. 644.}

The limitations of Einstein's principle-theory approach to SR have been noted by a number of commentators since 1905, including Wolfgang Pauli  and Arthur Eddington in the 20s, W. F. G. Swann in the 40s, and Lajos J\'{a}nossy and John S. Bell in the 70s, and Dennis Dieks in the 80s\footnote{See \citet{pauli81}; \citet{eddington28}, p. 7; \citet{swann41}; \citet{janossy71}; \citet{bell76,bell92}; and \citet{dieks84}.} All of these authors called for a more constructive version of SR. It was perhaps Bell who made the point in the clearest fashion.
\begin{quotation}
  If you are, for example, quite convinced of the second law of
  thermodynamics, of the increase of entropy, there are many things
  that you can get directly from the second law which are very
  difficult to get directly from a detailed study of the kinetic
  theory of gases, but you have no excuse for not looking at the
  kinetic theory of gases to see how the increase of entropy actually
  comes about. In the same way, although Einstein's theory of special
  relativity would lead you to expect the FitzGerald contraction, you
  are not excused from seeing how the detailed dynamics of the system
  also leads to the FitzGerald contraction.\footnote{\citet{bell92}.}
  
  \end{quotation}
  What is remarkable is that Bell himself seemed to be unaware of Einstein's own distinction between principle and constructive theory, and his repeated references to the analogy between SR and thermodynamics. At any rate, Bell stressed that he had no ``reservation whatever about the
power and precision of Einstein's approach''; his main point was that ``the
longer road  [a dynamical account of contraction and dilation] sometimes gives
more familiarity with the country''.\footnote{\citet{bell76}. For a discussion of Bell's 1976 treatment of SR by way of a ``Lorentzian pedagogy'', see \citet{brownpooley01} and \citet{brown05b}.}

\section{The CBH historical fable}

Let us return to the CBH argument. These authors offered a thought-provoking historical fable wherein SR began with Minkowski, who proposed a non-Newtonian geometry of space-time, and only later did Einstein come up with his principle theory approach. CHS regarded Minkowski as providing an ``algorithm for relativistic kinematics'', presumably based on the group of isometries of the postulated space-time structure, whereas in their fable they saw Einstein as furnishing an {\em interpretation} for SR: ``a description of the conditions under which the [Minkowski] theory would be true, in terms of certain principles that constrain the law-like behaviour of physical systems''. Analogously, it was argued, the CBH theorem could be viewed as providing an interpretation of quantum theory, based on information-theoretic constraints. It is clear from the CBH article that the authors regarded such an interpretation as having much in common with a position widely attributed to Niels Bohr, to the effect that quantum mechanics is not about micro-physical reality {\em per se} but rather the way we talk about it.

In attempting to evaluate CBH's neo-Bohrian stance, it is worth recalling first that the dominant viewpoint in the philosophy of space-time physics over the last few decades puts a very different gloss on Minkowski's contribution to SR. Far from being the basis of a mere algorithm for SR, the current orthodoxy seems to be that Minkowskian geometry provides a {\em constructive} dimension to SR (though it is not always put in these terms), and thereby significantly enhances its explanatory power. According to this view, it is the structure of the Minkowski space-time in which they are immersed that ultimately explains why rods and clocks in motion contract and dilate respectively.\footnote{See \citet{balashovjanssen03} and \citet{brownpooley06}.} But it is also worth bearing in mind that this was not entirely Minkowski's own interpretation of the four-dimensional geometry that bears his name. Minkowski's original position was much more like Poincar\'{e}'s (who indeed by 1906 had anticipated core features of Minkowski's work). It was that the Lorentz coordinate transformations can be seen as orthogonal transformations preserving the metrical properties of space-time, but the physical significance of these transformations derives from the fact that they are elements of the newly-discovered, or rather postulated, covariance group of all the non-gravitational interactions. The geometry does not come first---it is the dynamical symmetries that are fundamental, and susceptible to geometrical codification.\footnote{See \citet{brown05b}, ch 8.} In short, on either of these two views of the significance of Minkowski's contribution,  it amounts to a great deal more than a mere algorithm.

It is arguable that Minkowski's own reasoning  is not at root incompatible with the currently unorthodox dynamical interpretation of relativistic kinematics outlined in the previous section. The starting point of this account is indeed the Lorentz covariance of the equations governing all the non-gravitational forces---which in turn account for the cohesive properties of rigid bodies and clocks. We are not dealing here with a fully-fledged constructive theory, because the full details of the quantum theory of such interactions (and quantum theory it must be) are not required in the story. But such a theory would go a long way to avoid Einstein's self-confessed ``sin'' of treating rods and clocks as structureless, primitive entities, and the treating of space-time intervals as entities of a special type in the explanatory scheme of things.

It is not our purpose here to defend this dynamical, semi-constructive approach to relativistic kinematics.\footnote{For such a defense, see \citet{brownpooley01,brownpooley06} and \citet{brown05b}.} It is rather to point out that Einstein's original 1905 formulation of SR has its limitations, as Einstein himself knew full well and did not seek to hide. It is far from clear that he would have encouraged the use of SR---his 1905 SR---as a template for an `interpretation' of quantum theory. Or rather, for a {\em fundamental} interpretation. It is a remarkable thing that what might be called the kinematic structure of quantum theory, the nature of its observables and state space structure, can it seems be given a principle-theory, or `thermodynamic' underpinning. As Bell stressed, the beauty of thermodynamics is in its economy of reason, but the insight it provides is limited in relation to the messier story told in statistical mechanics. 

In assessing the import of the CBH theorem, Jeffrey Bub wrote:

\begin{quotation}
\noindent Assuming the information-theoretic constrainsts are in fact satisfied in our world, no mechanical theory of quantum phenomena that includes an account of measurement interactions can be acceptable, and the appropriate aim of physics at the fundamental level becomes the representation and manipulation of information.\footnote{\citet{bub04}, p. 242.}
\end{quotation}

The reasoning behind this remarkable conclusion that no mechanical account of the measurement process in quantum mechanics is viable, seems at first sight to be the analogue of the argument in SR that because Einstein treated rods and clocks as primitive entities in 1905, no analysis of their behaviour  {\em qua} moving atomic configurations is appropriate. An argument flatly rejected by Einstein himself. 

However, it should be noted that a key part of Bub's 2004 argument is that the historical success of statistical mechanics, and in particular recognition that the molecular-kinetic theory is more than a `useful fiction', came about because of Einstein's theory of Brownian motion. This theory not only allowed molecules to be counted, but demonstrated the limits of validity of thermodynamics. Where, Bub effectively asks, is the analogue of such superiority of constructive thought---the analogue of fluctuation phenomena---in quantum mechanics?

\begin{quotation}
\noindent The methodological moral I draw from the thermodynamics case is simply that a mechanical theory that purports to solve the measurement problem is not acceptable if it can be shown that, {\em in principle}, the theory can have no excess empirical content over a quantum theory. By the CBH theorem, given the information-theoretic constraints any extension of a quantum theory, like Bohmian mechanics, must be empirically equivalent to a quantum theory, so no such theory can be acceptable as a deeper mechanical explanation of why quantum phenomena are such subject to the information-theoretic constraints. To be acceptable, a mechanical theory that includes an account of our measuring instruments as well as the quantum phenomena they reveal (and so purports to solve the measurement problem) {\em must violate one or more of the information-theoretic constraints}.\footnote{\citet{bub04}, p. 261.}
\end{quotation}

Yet it is very doubtful whether Einstein advocated recognition of boosted rods and clocks as ``moving atomic configurations'' in SR because he thought such a view might ultimately lead to a violation of one or more of this 1905 postulates. It is more plausible that he did so because it made sense conceptually.\footnote{It is however interesting to ask whether there actually is an analogue of Brownian motion in the dynamical interpretation of SR. A positive answer, which appeals to certain phenomena in quantum field theory such as the Scharnhorst effect, is defended in \citet{brown05b}, ch. 9.} Likewise, disillusionment with the crude instrumentalistic nature of key aspects of Bohr's philosophy is justifiably one of the motivations for alternative interpretations of quantum theory---whether they involve an ``extension'' to the quantum formalism (such as the de Broglie-Bohm trajectories, or the collapse mechanism of GRW-type theories) or not (such as the Everett interpretation).\footnote{For further arguments in this vein, in particular defending the de Broglie-Bohm theory from Bub's 2004 criticism, see \citet{timpson04}, pp. 218--222.}

\section{Acknowledgments}

We wish to thank Bill Demopoulos for the kind invitation to contribute to this volume in honour of Jeff Bub, and to applaud him for conceiving and undertaking this project. We feel privileged. For nearly four decades Jeff Bub has been a leading figure in the foundations of quantum mechanics, through work characterized by honesty, rigour and penetration. Long may it continue.

                       

\end{document}